\documentclass[fleqn,usenatbib]{mnras}

\usepackage{newtxtext,newtxmath}

\usepackage[T1]{fontenc}

\DeclareRobustCommand{\VAN}[3]{#2}
\let\VANthebibliography\thebibliography
\def\thebibliography{\DeclareRobustCommand{\VAN}[3]{##3}\VANthebibliography}

\usepackage{graphicx}	
\usepackage{amsmath}	
\usepackage{natbib}

\usepackage{amssymb}	

\newcommand{\emerlin}[1]{\emph{e}-MERLIN#1}
\newcommand{\emerge}[1]{\emph{e}-MERGE#1}
\newcommand{\microJybm}{\mathrm{\umu Jy\,beam^{-1}}}

\newcommand{\sn}{$\mathrm{S/N}$}

\title[Revisiting a flux recovery systematic error]{Revisiting a flux recovery systematic error arising from common deconvolution methods used in aperture-synthesis imaging}

\author[J.~F.~Radcliffe et al.]{Jack F. Radcliffe,$^{1,2,3}$\thanks{E-mail: jack.radcliffe@up.ac.za}
 R.~J.~Beswick,$^{2}$ A.~P.~Thomson,$^{2}$  A.~Njeri,$^{2,4}$ and T.~W.~B.~Muxlow$^{2}$ \\
$^{1}$Department of Physics, University of Pretoria, Lynnwood Road, Hatfield, Pretoria, 0083, South Africa\\
$^{2}$Jodrell Bank Centre for Astrophysics, University of Manchester, Oxford Road, Manchester M13 9PL, UK \\
$^{3}$National Institute for Theoretical and Computational Sciences (NITheCS) South Africa \\
$^{4}$School of Mathematics, Statistics and Physics, Newcastle University, Newcastle upon Tyne, NE1 7RU, UK 
}

\date{Accepted 2023 August 31. Received 2023 August 31; in original form 2022 November 28}

\pubyear{2023}

\begin{document}
\label{firstpage}
\pagerange{\pageref{firstpage}--\pageref{lastpage}}
\maketitle

\begin{abstract}
The point-spread function (PSF) is a fundamental property of any astronomical instrument. In interferometers, differing array configurations combined with their $uv$ coverage, and various weighting schemes can produce an irregular but deterministic PSF. As a result, the PSF is often deconvolved using CLEAN-style algorithms to improve image fidelity. In this paper, we revisit a significant effect that causes the flux densities measured with any interferometer to be systematically offset from the true values. Using a suite of carefully controlled simulations, we show that the systematic offset originates from a mismatch in the units of the image produced by these CLEAN-style algorithms. We illustrate that this systematic error can be significant, ranging from a few to tens of per cent. Accounting for this effect is important for current and future interferometric arrays, such as MeerKAT, LOFAR and the SKA, whose core-dominated configuration naturally causes an irregular PSF. We show that this offset is independent of other systematics, and can worsen due to some factors such as the goodness of the fit to the PSF, the deconvolution depth, and the signal-to-noise of the source. Finally, we present several methods that can reduce this effect to just a few per cent.
\end{abstract}

\begin{keywords}
techniques: interferometric, photometric -- methods: data analysis
\end{keywords}



\section{Introduction}\label{s:intro}

The point spread function (PSF) is a fundamental property of any astronomical instrument and describes the response of an instrument to a normalised point source. It characterises how much the true sky brightness distribution is distorted by the instrument and is often used to define the angular resolution of the instrument. Understanding and removing the PSF contribution is of critical importance in extracting accurate photometry from your image. For single aperture instruments, the PSF shape is dependent on many factors such as obscuration, aberrations, and pointing errors. This is exacerbated for ground-based instruments where thermal and gravitational effects come into play. As a result, the PSF is often time-varying and thus sophisticated modelling is required to characterise it properly \citep[e.g.,][]{krist2011}. However, single-aperture instruments have one key advantage and that is a filled aperture. This means that the resultant PSF is often well-behaved, with most of its amplitude located in a main directional lobe and, with minimal side lobes that only affect high dynamic range imaging.

In contrast, the PSF of an interferometer (often known as the `dirty beam' or `synthesised beam') does not have this luxury, and the incomplete Fourier sampling of the aperture means that the PSF often has significant sidelobe structure that can affect the extraction of accurate photometry. However, the PSF of an interferometer is highly deterministic and can be calculated exactly from the visibilities measured by the interferometer. This means that interferometric images can often be improved by deconvolving the PSF from the image. The most commonly used method of deconvolution is the CLEAN family of algorithms. These algorithms deconvolve through a process of iterative peak finding and PSF subtraction, down to some predetermined threshold (also known as `matching pursuit'). However, the estimated sky brightness model obtained from these algorithms is typically comprised of delta functions and/or Gaussians which can often look distinctly unphysical. As a result, a `restoration' step is usually employed. In this step, the model is convolved with a 2D Gaussian fitted to the PSF that estimates the effective resolution of the interferometer as if it had a filled aperture. This is then added to the subtracted image (which can contain low-level unresolved flux) and this restored image is then used for science.

However, this restoration step can result in a systematic error that occurs when extracting the flux densities of sources. This was noted early when the Fourier sampling was sparse \citep[e.g.,][]{hogbom1974}. As a result, early array designs were often optimised to generate an approximately Gaussian PSF (e.g., the Very Large Array - VLA). The error was first quantified by \citet{1995Jorsater} who found that there was a mismatch between the flux density of an input model source and the resultant recovered flux densities when CLEAN deconvolution was used. This was found to be from contamination by faint flux that had not been deconvolved. While this effect has been routinely corrected for in spectral line studies \citep[e.g.,][]{Kennicutt2007,cannon2009,hunter2012,deblok2018,novak2020}, it had sparingly been taken into account during continuum imaging studies \citep[e.g.,][]{benisty2021,Heywood2022,booth2023}.

This is understandable, as the effect of this systematic error had been lessened through the improved $uv$ coverage and array designs that resulted in PSFs that more closely resemble a filled aperture. However, the designs of modern interferometric arrays, such as MeerKAT, ASKAP, LOFAR, ALMA, the upcoming Square Kilometre Array (SKA) and next-generation VLA (ngVLA), are driven by multiple, diverse scientific objectives that require a high sensitivity to diffuse structures, and high angular resolutions, and non-imaging applications, such as pulsar timing, all at the same time. These competing scientific requirements have resulted in the configuration of these arrays being a compromise, resulting in the majority being core-dominated with a sparser distribution of antennas on longer baselines. As a result, the large number of shorter baselines produces a naturally irregular PSF. 

It is in this context that we are re-examining this systematic effect as will be important to account for when using these modern interferometric arrays (as long as the use of CLEAN-style deconvolution methods persists). In this paper, we shall use a controlled simulation of a core-dominated radio interferometer array to quantify this effect on recovered flux densities, and present some methods that can help correct it so that interferometric surveys can take this systematic error into account.

The paper is organised as follows. In Section~\ref{s:sims}, we introduce the simulations used. Our results are shown in Section~\ref{s:res}, and we explain the origin of the offsets in Section~\ref{s:offsets}. We present the methods to solve for the effects of the PSF in Section~\ref{s:solns}, and discuss the implications and limitations of these results in Section~\ref{s:conclusions}.

Throughout this paper, we define the spectral index, $\alpha$, as $S_\nu \propto \nu^{\alpha}$ where $S_\nu$ is the specific integrated flux density (per unit frequency, $\nu$). We use the convention used in the Common Astronomy Software Applications (CASA) package \citep{casa2022} for the robustness factor ($R$) of the Brigg's weighting scheme \citep{briggs1999} which is defined to be between $-2$ and $2$. These values approach uniform and natural weighting, respectively.  
 
\section{Simulations}\label{s:sims}

For the simulations, we opted to generate a combined VLA (in A-configuration) and \emerlin{} array. This combination has a well-documented irregular PSF due to the short spacings of the VLA compared to the \emerlin{} array \citep[e.g., see][]{muxlow2020}. We also have generated a representative MeerKAT array to prove that this effect is evident in modern core-dominated interferometers (see Section~\ref{s:conclusions}). 

To build the simulated data, \textsc{simms}\footnote{\href{https://github.com/ratt-ru/simms}{https://github.com/ratt-ru/simms}} was used to create empty CASA measurement sets with a delay-tracking centre located at a Right Ascension of $12^\mathrm{h}$ and a Declination of $+60^\circ$ (or $-30^\circ$ for MeerKAT), and a total on-source integration time of 12~hours. These data sets have four correlations present with circular feeds for VLA and \emerlin{}, and linear feeds for MeerKAT. We used a bandwidth representative of a standard L-band ($\sim 1.5\,\mathrm{GHz}$) observation with each array. The long integration times (10\,s) and a small number of frequency channels were used to keep the data sizes as small as possible without adversely changing the PSF profile. The exact specifications of each simulated data set are summarised in Tables~\ref{tab:properties_sims} and~\ref{tab:SEFD_emerlin}.  

Thermal noise was inserted into the visibilities for each baseline, $\mathsf{V}_{pq}$, that comprises antennas $p$ and $q$. We assumed that the system thermal noise was broadband and so can be described using a circularly complex Gaussian probability distribution $\mathcal{N}$ that has a zero mean and a variance of $\sigma^2_{pq}$ per visibility. The standard deviation of the visibility scatter, $\sigma_{pq}$, is given by the radiometer equation that, per visibility, is given by,
\begin{equation}
 	\sigma_{pq} = \frac{1}{\eta_c}\left(\frac{{\rm SEFD}_p \times {\rm SEFD}_q}{2t_\mathrm{int}\delta\nu}\right)^{\frac{1}{2}},
\end{equation}
where SEFD is the antenna's System Equivalent Flux Density, $\eta_c$ is the efficiency factor (taken to be $0.88$ assuming 4-level quantisation), $t_\mathrm{int}$ is the integration time, and $\delta\nu$ is the channel bandwidth. This was then added into the visibilities of the measurement set as an additive term per polarisation,
\begin{equation}
	\mathsf{V}^{\prime}_{pq} = \mathsf{V}_{pq} + \mathcal{N}(0,\sigma^2_{pq}).
\end{equation}
To ensure that the natural weighting scheme generates an image with the optimal sensitivity, the weights of each visibility, $W_{pq}$, were set such that, 
\begin{equation}
	W_{pq} = \frac{1}{\sigma_{pq}^2}.
\end{equation}
This is important for the heterogeneous \emerlin{} array where the majority of the sensitivity is provided by the baselines to the more sensitive Lovell and Cambridge telescopes. These visibilities should now be representative of the thermal noise of the array given their nominal antenna sensitivities. Next, elevation-dependent flagging was conducted by removing data from each antenna when the pointing centre is below $5$ degrees elevation for that antenna. 

\begin{table}
	\centering
	\caption{The properties of the simulated data sets. The sensitivities were calculated using a natural weighting scheme when imaging.}
	\label{tab:properties_sims}
	\begin{tabular}{ccccc}
		\hline
		Array      & Frequency    & SEFD                             & \# channels & Sensitivity    \\
		           &   (GHz)      & (Jy)                             &             & ($\microJybm$) \\
		\hline
		VLA        & 1.024--2.048 & 420                              & 32          & 1.91 \\
		\emerlin{} & 1.254--1.766 & see Table~\ref{tab:SEFD_emerlin} & 64          & 6.04 \\
		MeerKAT    & 0.900--1.677 & 450                              & 32          & 1.15\\
		\hline
	\end{tabular}
\end{table}
\begin{table}
	\centering
	\caption{System equivalent flux densities used for the \emerlin{} antennas.}
	\label{tab:SEFD_emerlin}
	\begin{tabular}{cc} 
		\hline
		Antenna   & SEFD \\
		          & (Jy) \\
		\hline
		Lovell    & 40 \\
		Knockin   & 400 \\
		Pickmere  & 450 \\
		Darnhall  & 450 \\
		Defford   & 350 \\
		Cambridge & 175 \\
		\hline
	\end{tabular}
\end{table}

With the noise inserted into the data, we then generated the sky model. To keep the number of variables as small as possible, our input sky model was comprised of a total of 13,500 point sources. This was split into three individual realisations each with 4,500 sources. Point sources were selected as our model to eliminate any complications arising from the `resolving out' of flux due to the incomplete Fourier sampling of an interferometric array. These sources had randomly assigned integrated flux densities which were designed to cover a signal-to-noise (\sn{}) range of 5--70. The sources were distributed randomly with the caveat that each source was at least $10\arcsec$ from its nearest neighbour. This was chosen to eliminate any complications arising from source confusion when attempting to measure the individual flux densities. In addition, the point sources were defined to be located at an integer number of pixels in the model image to prevent any pixelation errors when it was transformed into model visibilities.

This sky model was inserted into the measurement sets by converting the model sky (via de-gridding) into model visibilities using the \textsc{wsclean} \citep{offringa2014} package. This method ensures that wide-field effects, such as baseline non-coplanarity (known as the $w$-term), are not present in the modelled sky. This model was then added into the visibilities using the CASA task \textsc{uvsub}. For the subsequent imaging runs, \textsc{wsclean} was used due to its versatility and speed. We cross-checked the imaging outputs of \textsc{wsclean} and CASA's \textsc{tclean} task when using identical parameters and we found an average fractional difference of just 0.3\%. We are therefore confident that our results are minimally impacted by the choice of imaging software. The exact imaging runs are presented in Section~\ref{s:res}.

\section{Results}\label{s:res}

\begin{figure*}
	\includegraphics[width=\linewidth]{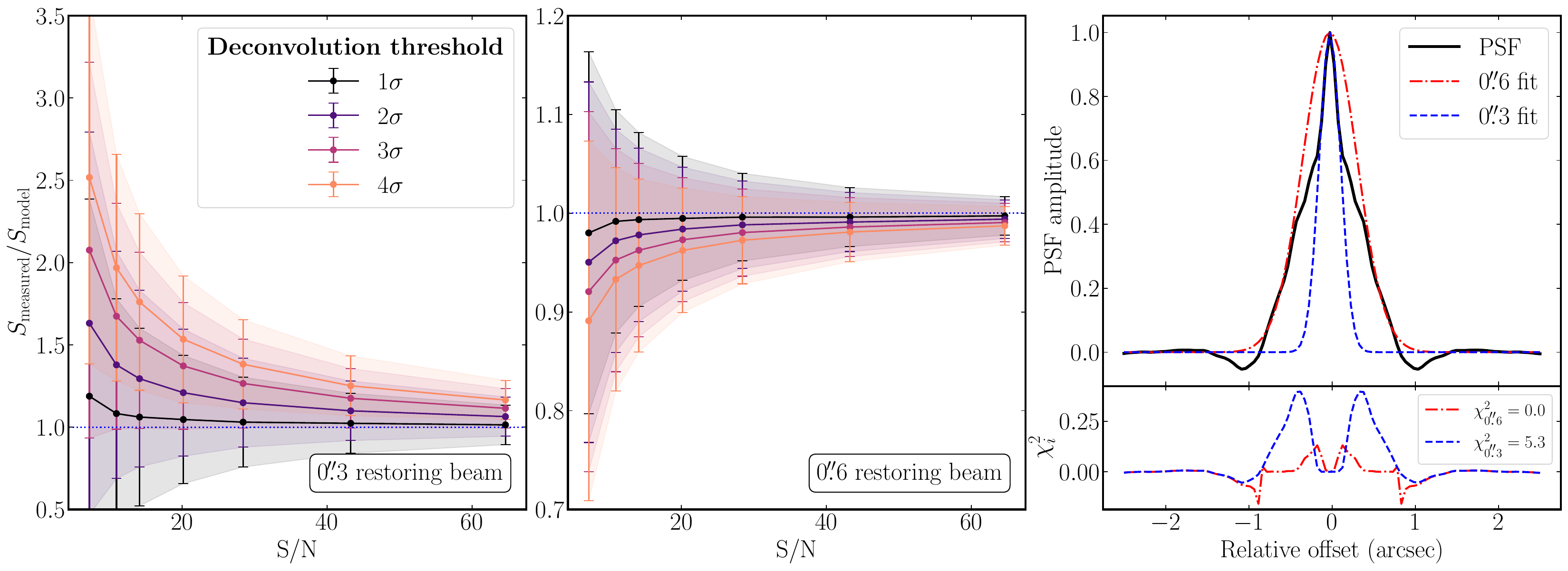}
	\caption{\textit{Left and centre panels:} The flux density recovery performance of the simulated combined VLA and \emerlin{} arrays using two different Gaussian to the PSF. The binned \sn{} of the 13,500 sources are plotted against the ratio of the measured to model flux densities for a range of deconvolution thresholds. The error bars and shaded regions correspond to $1\sigma$ deviations within the bins. The fits are $\sim0\farcs3$ (left panel) and $\sim0\farcs6$ (centre panel) which correspond to the resultant fits by CASA's \textsc{tclean} task and the stand-alone \textsc{wsclean} package. \textit{Right panel:} A 1D slice through the PSF with the resultant fits overlaid. The bottom panel shows the residuals are plotted per pixel $\left(\chi^2_i = \mathrm{(PSF - fit)^2/PSF}\right)$ and the $\chi^2 = \sum\chi_i^2$ values are presented.}
	\label{fig:PSF_fits}
\end{figure*}

To show the effects of an irregular PSF upon the resultant measured flux densities, the combined VLA and \emerlin{} array were imaged using a robust value of zero. This produces an irregular PSF with a secondary component on larger angular scales that is caused by the short spacings provided by the VLA. This is shown in the right-hand panel of Figure~\ref{fig:PSF_fits} which shows a 1D slice through the PSF.  Each image was $25\mathrm{k} \times 25\mathrm{k}$ pixels in size with each pixel corresponding to an angular size of $0\farcs05$. These choices enable all 4,500 sources of each realisation to be imaged at once. The PSF was deconvolved from the source brightness using the Clark algorithm \citep{clark1980}. We used multiple different deconvolution thresholds (i.e., the flux density level where deconvolution stops) ranging from 1--4 times the local root-mean-square (rms) noise levels ($\sigma$) so we can illustrate how the deconvolution process can drastically affect the recovered fluxes. Once deconvolution had ceased, the source brightnesses were restored by convolving the model with two different 2D Gaussian fits to the PSF; a $\sim 0\farcs3$ full-width half-maximum (FWHM) fit that was produced CASA's \textsc{tclean} task and a $\sim0\farcs6$ FWHM fit from \textsc{wsclean}. The fact that two different imaging packages produce completely different fits is indicative of the difficulty of assuming a Gaussian fit to a highly non-Gaussian PSF. The convolved model is then added back to the residual image (where the source flux densities have been removed and deconvolved). To summarise, we produced a total of eight images, with four different deconvolution thresholds, and two different fits to the PSF.

Flux density extraction was conducted using aperture photometry \citep[implemented using \textsc{photutils};][]{photutils}. We used an aperture that is 2.5 times larger than the major axis of the fitted PSF. This enables us to recover $>99\%$ of the source's flux density while compensating for the widening of the source structure due to the convolution of the model sky with the restoring beam. By using aperture photometry and the knowledge of where our sources are, we can mitigate the well-known effect of `flux-boosting' \citep[e.g.,][]{Jauncey1968}. This occurs because typical source-detection routines use a sigma detection threshold to determine what is a true source above the noise. In the low \sn{} regime, local noise variations can cause some sources to be co-located with a noise peak which results in the source exceeding the detection threshold, while those sources on a negative noise peak would not be detected. This will bias the distribution of sources upward in flux density in the low \sn{} regime. In addition, the fitting routines often employed perform inadequately in the low \sn{} regime \citep[e.g.,][]{hale2019}. Aperture photometry will not suffer from this issue as the noise contribution should (statistically) average to zero, while all the real flux located within the aperture will be detected. In addition, no sources will be missed (e.g., due to being co-located on a negative noise peak) as we know the source positions beforehand. 

Figure~\ref{fig:PSF_fits} shows the results of the flux extraction for the eight images described earlier. In the left and centre panels, we present the ratio of measured flux densities ($S_\mathrm{measured}$) to the model flux densities ($S_\mathrm{model}$) against the \sn{} ratio for a range of deconvolution thresholds. The flux densities of the sources are median binned into 7 bins with each containing approximately an equal number of sources. The error bars and filled-in areas correspond to the $1\sigma$ standard deviation of the flux densities within each bin. In the right panel, we show a 1D slice of the PSF with the corresponding Gaussian fit which is used to produce the output images.

We would expect that the ratio of $S_\mathrm{measured}/S_\mathrm{model}$ would be approximately equal to one if the flux densities were recovered perfectly. However, as we can see in the left and centre panels of Figure~\ref{fig:PSF_fits}, the flux densities recovered differ significantly from those inputted. There are four main results from this that we need to draw attention to. 

Firstly, regardless of the fit, the recovered flux densities are different to the model flux densities. These offsets can be significant and can range from just 5\% to 250\%. Such a systematic error can surpass the typical amplitude errors of interferometers which are often taken to be of the order of 10\%.  Secondly, the extent of the flux density offsets worsens when we deconvolve to higher thresholds (or a `shallower' deconvolution). This indicates that the flux densities recovered in any interferometric array are highly dependent on the deconvolution depth. For interferometric arrays with a large number of elements (e.g., LOFAR, MeerKAT, SKA), the additional computational requirements to deconvolve to much lower flux densities could be prohibitive, subsequently adversely affecting the measured flux densities of these arrays. 

Thirdly, the flux density offset becomes more severe towards the low \sn{} regime showing that flux uncertainties at the noise limit are increasingly uncertain irrespective of the increasing fitting errors. The PSF must be taken into account within this regime. Finally, the flux density offsets are more extreme when the fit between the restoring beam is worse. In the next section, we shall explain how the flux density systematics arise and their relationship to both the \sn{} and deconvolution.

\section{The origin of the measured flux density errors}\label{s:offsets}

\begin{figure*}
	\includegraphics[width=\linewidth]{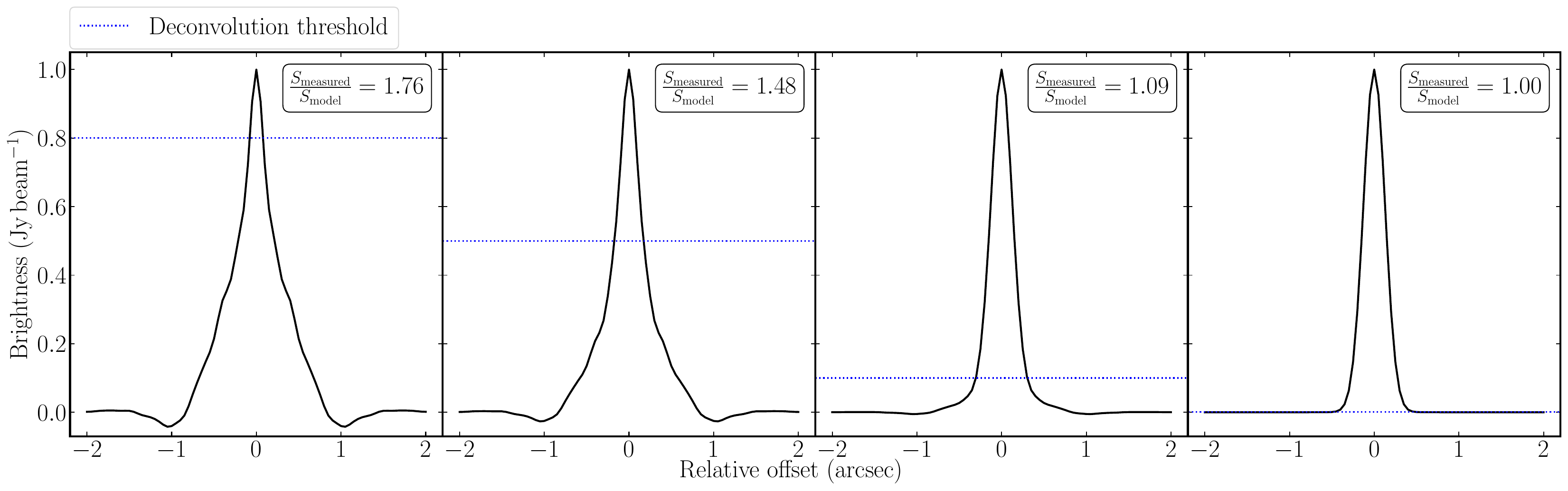}
	\caption{A 1D toy model of the deconvolution process using the PSF profile presented in Figure~\ref{fig:PSF_fits}. The deconvolution threshold is shown by the blue dotted line. As deconvolution proceeds to lower thresholds (from the left to the right panel), the number of beam widths which the flux density is measured across tends to be one. Theoretically, we would get a perfect recovery of the flux density if we were able to deconvolve to a 0\,Jy threshold.}
	\label{fig:clean_example}
\end{figure*}

As inferred previously, these trends are all related to the deconvolution method used. To illustrate this, we are going to use a toy model of the deconvolution process. For simplicity, we are going to assume that there is no thermal noise present and that we are observing a simple 1\,Jy point source. This source is observed with an instrument that has the same irregular PSF profile as shown in the right panel of Figure~\ref{fig:PSF_fits}. As expected, the measured sky brightness is just equal to the PSF response. The key here is that the interferometer measures the source brightness in units of Jansky per PSF.

The PSF contribution is then deconvolved from the true source brightness using the \citet{hogbom1974} CLEAN algorithm. In this method and many other deconvolution methods used in radio interferometry, the pixel of peak brightness is found and then a percentage of this brightness is convolved with the PSF and then subtracted from the image. The value and position of the brightness that is removed are then stored as a sky brightness model (as delta functions in the case of the \citet{hogbom1974} algorithm). This process is then re-iterated until a stopping criterion is met. Often this criterion is a threshold where deconvolution stops when the brightest pixel in the subtracted image falls below a pre-defined value. This threshold is often a few $\sigma$ above the noise for multiple reasons such as reducing computational expenditure, preventing the divergence of the deconvolution process, ensuring that a reliable sky model is generated (e.g., for self-calibration), or preventing unnecessary over-deconvolution which can lead to other systematics such as CLEAN bias \citep[e.g.,][]{becker1995}.

The model sky brightness is then convolved with a restoring beam which comes from a Gaussian fit to the PSF and is added back to the residuals (the image after the PSF multiplied by the brightness is subtracted). A Gaussian is often chosen as it is a good approximation for the idealised response of the interferometer as if it has complete $uv$ coverage. The restored image is assumed to have units of Jansky per beam where the beam is defined by the angular size of the Gaussian. A flux density is then measured by summing the brightnesses over the number of beams that the source brightness extends over.

For example, in our toy model, the PSF is fitted with a $\sim 0\farcs3$ Gaussian that is used as the restoring beam. This gives us a fixed solid angle which we can integrate over to obtain a flux density. In Figure~\ref{fig:clean_example}, we show the deconvolution process at four progressively lower thresholds. At the lowest threshold (0.001\,Jy; first panel from the right in Fig.~\ref{fig:clean_example}), the brightness in the output image covers almost a single beam area therefore we recover the input flux density exactly. However, when the PSF is still present in the output image (as in the other three panels), the brightness of the source is summed over more than one beam area hence a higher flux density is measured when compared to the input model. This explains why the 0\farcs6 fit, which is larger than the true PSF, causes a reverse effect where the measured flux density is smaller than the input model flux density (as shown in the central panel of Fig.~\ref{fig:PSF_fits}). In this case, there are fewer beam elements that the brightness is observed over, hence the recovered flux density is measured to be lower than the model. 

In other words, above the deconvolution threshold, the recovered sky brightness has units of $\mathrm{Jy\,beam^{-1}}$, but below the deconvolution threshold, the flux density is still in units of $\mathrm{Jy\,PSF^{-1}}$. This means that the source of this error occurs when the residuals (which has units of $\mathrm{Jy\,PSF^{-1}}$) are added back to the convolved model (which has units of $\mathrm{Jy\,beam^{-1}}$), but we extract the flux densities from the image assuming units of $\mathrm{Jy\,beam^{-1}}$ throughout. The subsequent mismatch between PSF and beam area below the deconvolution threshold, therefore, causes the flux density to be miscalculated. These flux density offsets get progressively worse as the deconvolution threshold increases, as is shown in Figure~\ref{fig:clean_example}. This is because the percentage of flux restored with the fitted beam width/solid angle (i.e., converted from units of $\mathrm{Jy\,PSF^{-1}}$ to $\mathrm{Jy\,beam^{-1}}$) is increased when using a lower threshold. This argument also explains the trends with \sn{} seen in Figure~\ref{fig:PSF_fits}. When a fixed deconvolution threshold level is used, sources with a lower \sn{} have less of their total flux density deconvolved and restored with the fitted beam which has a well-defined solid angle. This results in a larger proportion of the flux density being incorrectly estimated once measured. 

\section{Possible solutions}\label{s:solns}

\begin{figure*}
	\includegraphics[width=\linewidth]{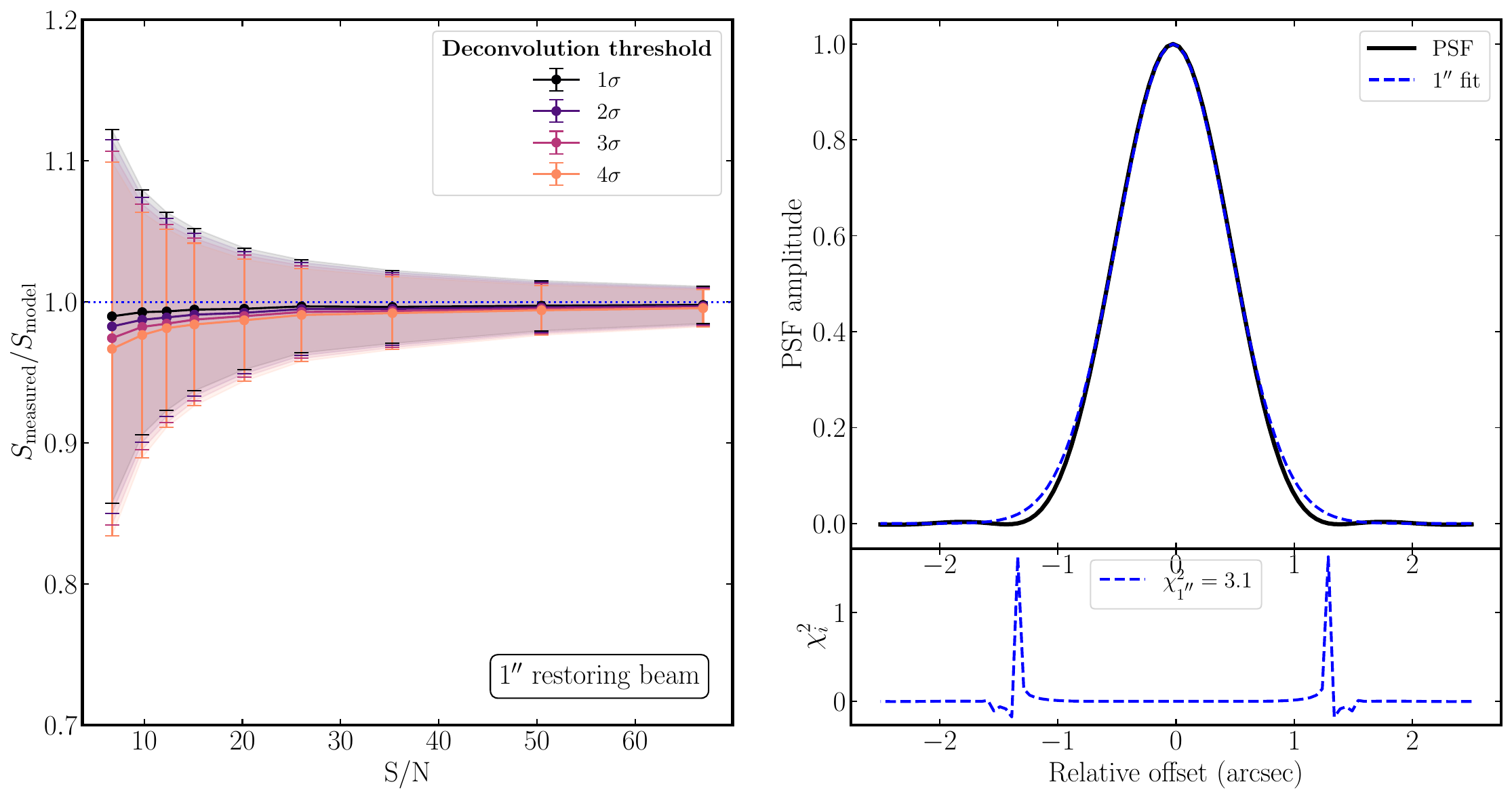}
	\caption{\textit{Left panel:} The flux density recovery performance of the simulated combined VLA and \emerlin{} arrays using a reweighing scheme that forces the resultant PSF to closely represent a Gaussian with FWHM of 1\arcsec. The \sn{} of the 4500 sources is plotted against the binned ratio of the measured to model flux densities for a range of deconvolution thresholds. The vastly improved performance compared to Figure~\ref{fig:PSF_fits} is evident and the flux density offsets are reduced to just a few per cent for the faintest sources. \textit{Right panel:} A 1D slice through the PSF with the resultant fit overlaid. The bottom panel shows the residuals with the same definition as Figure~\ref{fig:PSF_fits}. The slight overestimation of the PSF size causes a slight underestimate of the measured flux densities.}
	\label{fig:gaus_fits}
\end{figure*}

Now that the issue has been outlined and explained, we shall introduce some possible methods that can reduce these flux density offsets. These solutions are not only motivated by the simulations presented but also through experiences with real data. 

\subsection{Making the PSF closer to a Gaussian through re-weighting}

The motivation for this solution comes from the \emerge{} survey \citep{muxlow2020}. This survey uses a combined VLA and \emerlin{} array which results in a highly non-Gaussian PSF which caused flux density offsets in these data.\footnote{In particular, see Section~2.5.1 of \citet{muxlow2020} where this is discussed.} As part of this survey, solutions to this problem were formed and implemented.

There are three main ways in which the flux density offsets can be reduced. The first method is to deconvolve to deeper thresholds. This is shown in the left and centre panels of Figure~\ref{fig:PSF_fits}. However, it is worth noting that this may not be possible for real data. Firstly, deconvolution cannot continue indefinitely as the computational expense can become prohibitive but also, for many modern sensitive arrays, such as MeerKAT, it is often residual calibration errors (such as pointing errors) and direction-dependent effects that limit to which deconvolution can be performed.

It is therefore practically impossible to deconvolve all flux (especially that below the noise level) and fully remove this effect. The extent of this systematic could be removed in statistical studies (e.g., source counts) using simulations as the exact extent of the flux density offsets will be sensitive to any process that changes the PSF shape. This would include factors such as the exact $uv$ coverage,\footnote{This is in itself dependent upon a multitude of factors including the number of antennas and their location, source location, flagging of data and more.} and data weighting. However, on a source-by-source basis, extra care would still need to be taken to account for this systematic effect.

The second method would be to re-weigh the data so that the resultant PSF more closely represents a Gaussian. This means that the systematic effect is reduced and prevents complications caused by deconvolving too deeply to be mitigated. We achieved this with the simulated data sets using a Gaussian taper of 1 arcsecond. The results are shown in Figure~\ref{fig:gaus_fits}. As shown, the flux density offsets are still present (due to the imperfect Gaussian PSF), but to a lesser extent due to the much better fit. The offsets are only a few per cent compared to Figure~\ref{fig:PSF_fits} which could be more than 5--10\%. In addition, the spread between the various deconvolution thresholds has been reduced with differences of around $\sim1\%$ on the lowest \sn{} sources. However, as with any re-weighting scheme, there are always sacrifices that have to be made whether it be resolution or point-source sensitivity. In this case, the sensitivity of the observation worsened by 90\% which would be detrimental to many surveys. In such cases, a compromise should be required and so would need to subsume this systematic effect into their error budget for individual sources. 

This was the approach that was used for the \emerge{} survey \citep{muxlow2020}. We found for core-dominated arrays, as that previously encountered in \emerge{}, applying a Tukey taper \citep{tukey62} allowed us to make the PSF more closely represent a Gaussian, and reduce the systematic flux offsets to just a few per cent. The taper allowed us to smoothly reduce the re-weight of the shorter baselines which in turn, reduces the amplitude of the shoulders in the PSF. While the sensitivity was degraded by around $30\%$ the resultant flux densities were now reliable when compared to other surveys.

Unfortunately, the sheer number of different weighting schemes for interferometric data means that it should be viable to obtain these compromises, but the exact parameters are often difficult to determine analytically as the resultant PSF shape is dependent upon many interconnected factors. This means that a Monte-Carlo approach to the exact compromise would often be required for each observation.

\subsection{Rescaling the residuals}\label{ss:im_strat}

\begin{figure}
	\includegraphics[width=\linewidth]{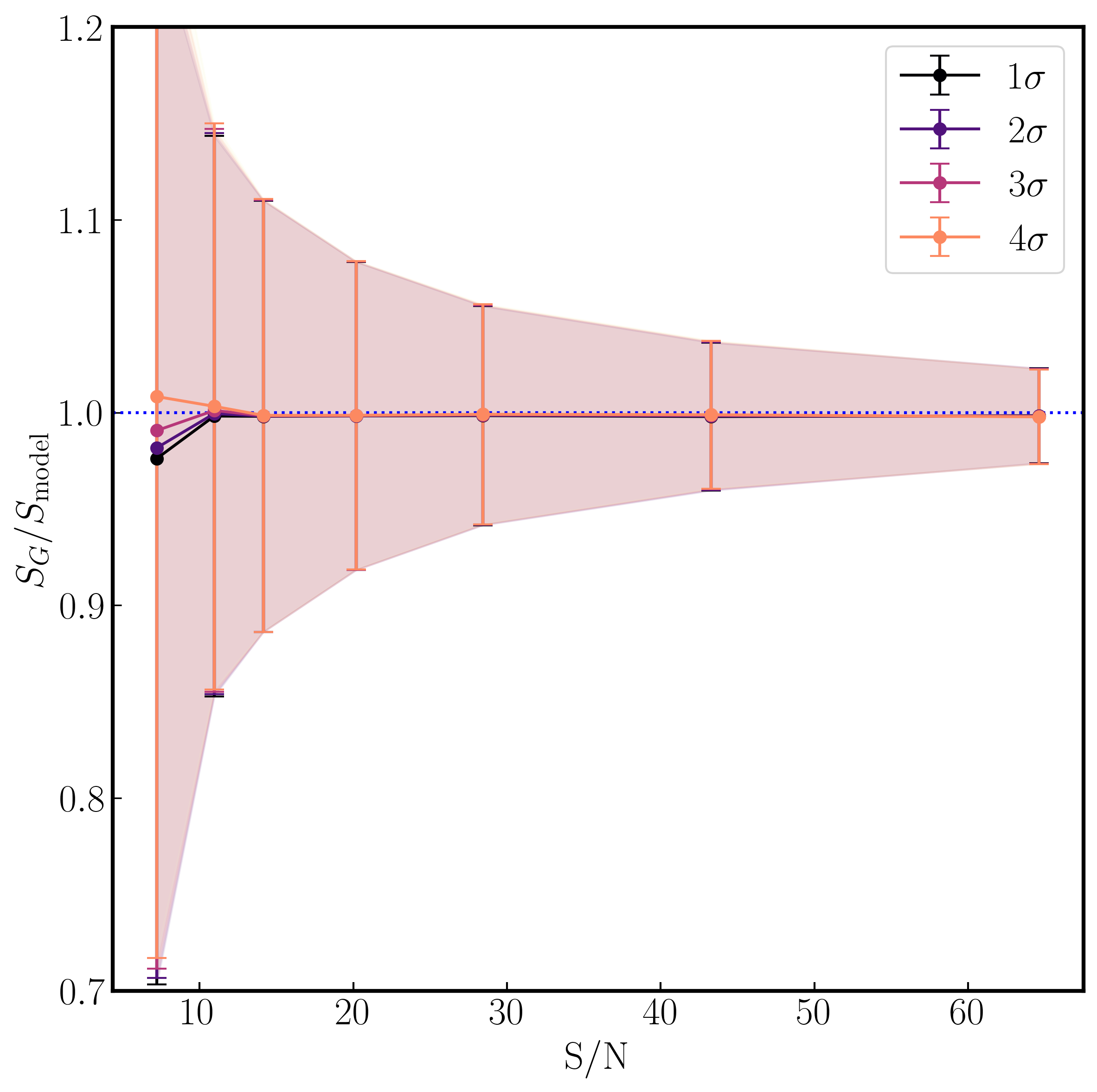}
	\caption{The estimated true flux density $S_G$ achieved  by rescaling the dirty map using Equation~\ref{eq:jor}. The restoring beam was chosen to be of 0\farcs6 as in the centre panel of Figure~\ref{fig:clean_example}.}
	\label{fig:jorsater}
\end{figure}

An alternative method was proposed by \citet{1995Jorsater} (see their Appendix A.2) and is the method commonly used in many spectral line studies. Here, the mismatch between the restoring beam and PSF areas, which causes the flux density systematic error, can be estimated and corrected by rescaling the measured flux densities to be in units of per beam rather than per PSF. Their formulation is as follows. If we take a residual flux density, $R$, and a restored/deconvolved flux density, $C$, the unknown true source flux density, $G$, is given by residual map, $R$, rescaled by a factor $\epsilon$ by,
\begin{equation}
	G = C + \epsilon R. \label{e:7}
\end{equation}
The factor $\epsilon$ is unknown and is required for us to be able to rescale the residual flux from units of $\mathrm{Jy\,PSF^{-1}}$ to $\mathrm{Jy\,beam^{-1}}$. This factor is related to the ratio of the areas through,
\begin{equation}
	\epsilon = \frac{\mathrm{Restoring~beam~area}}{\mathrm{PSF~area}}.
\end{equation}
To estimate $\epsilon$, \citet{1995Jorsater} suggested that you can conduct two separate deconvolution runs to different thresholds. This will result in two recovered fluxes, $C_1$ and $C_2$, and two residual fluxes, $R_1$ and $R_2$. These can be related to the true flux density, $G$, by the following two equations,
\begin{equation}
	G = C_1 + \epsilon R_1\quad\mathrm{and}\quad G = C_2 + \epsilon R_2.
\end{equation}
These can be solved for the two unknowns, $G$ and $\epsilon$ to give,
\begin{align}
	\begin{split}
		\epsilon &= \frac{C_2 - C_1}{R_1 - R_2}, \\
		G &= \frac{C_1R_2 - R_1C_2}{R_2-R_1}. \label{e:11}
	\end{split}
\end{align}
This can be simplified by using a single deconvolution run and the un-deconvolved or dirty map so $C_2 = 0$. This reduces Eqn.~\ref{e:11} to,
\begin{equation}
	\epsilon = \frac{C_1}{R_2-R_1} \quad \mathrm{and} \quad G = \epsilon R_2. \label{eq:jor}
\end{equation}
We applied Eqn.~\ref{eq:jor} to the imaging runs that used a 0\farcs6 restoring beam (see the centre panel of Figure~\ref{fig:PSF_fits}). We found that the true flux densities are almost perfectly recovered for all deconvolution thresholds at S/N larger than 15, far exceeding the performance of the re-weighting method presented in Section~\ref{ss:im_strat} (see Figure~\ref{fig:jorsater}). However, at low S/N, the recovery deviates with a few per cent errors on all deconvolution depths. In addition, the calculation of $\epsilon$ for all sources increases the scatter in an individual measurement whereas the binned statistical population is correct. 

An alternative method for correcting for this effect comes from the MeerKAT MeerKAT International Gigahertz Tiered Extragalactic Explorations (MIGHTEE) survey \citep{Heywood2022}. In this paper, they deal with the PSF mismatch by using a PSF homogenisation kernel which is convolved with the residual image (post-deconvolution model removal) to make the residuals match the restoring beam sizes. The model is then convolved with the restoring beam and added to the rescaled residuals. This ensures that the whole image is now in the same units. This method has the distinct advantage in removing the need to reweigh the data to get Gaussian beams but the open question still arises as to whether this artificially affects the Fourier scales that are expressed in the image. We leave this investigation to a future paper that will look at resolved sources. 

\begin{figure*}
	\includegraphics[width=\linewidth]{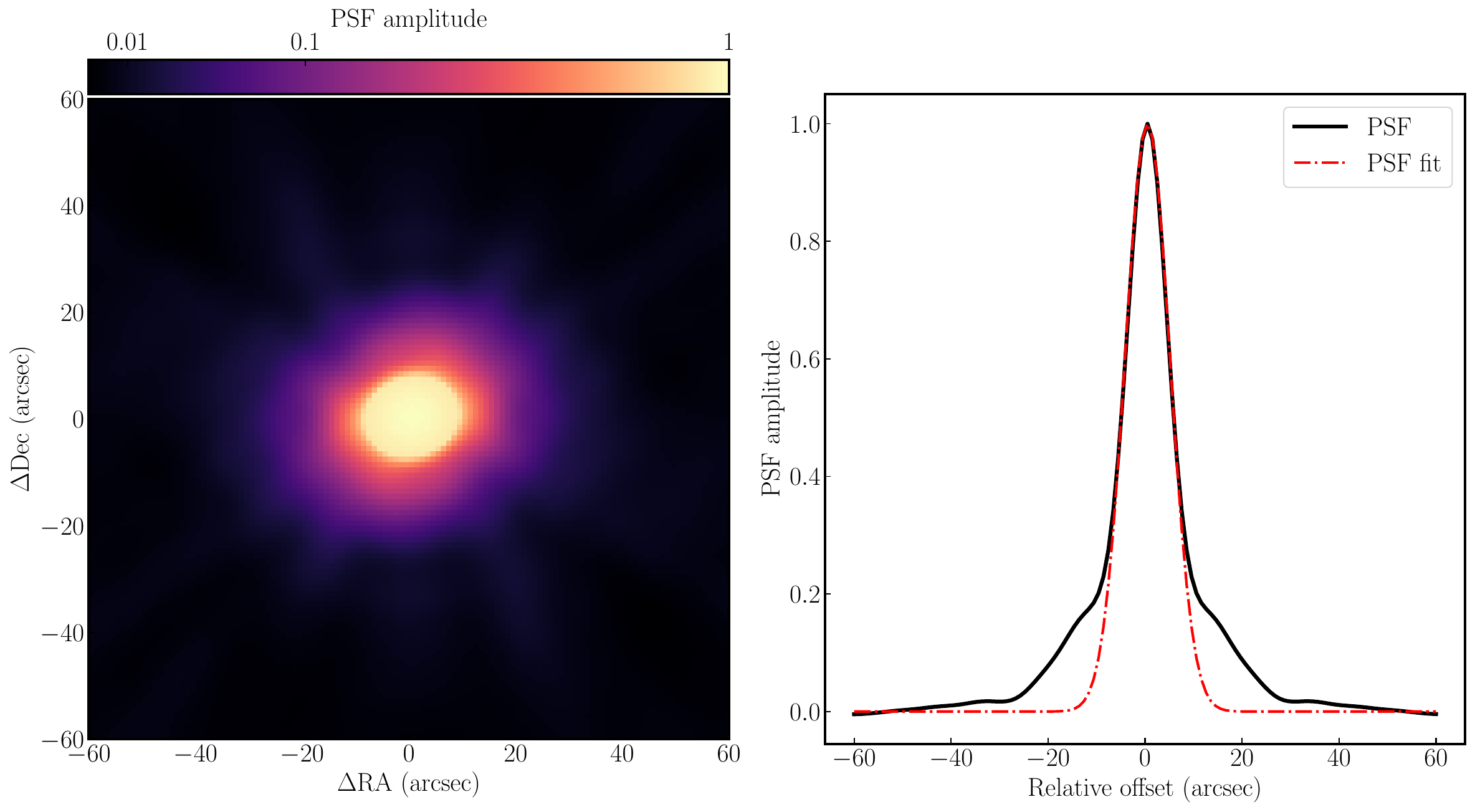}
	\caption{The PSF of a simulated 12\,hr MeerKAT observation of a source at $\delta = -30^\circ$ using robust weighting of 0.5. The left panel is a $1\arcmin \times 1\arcmin$ cutout of the PSF while the right panel is a 1D slice at $\Delta\mathrm{Dec}=0$ with the Gaussian fit to the PSF over-plotted. The shoulder of the PSF due to the large number of short baselines present in both panels which results in a Gaussian unable to fit the PSF.}
	\label{fig:meerkat_example}
\end{figure*}

\section{Discussion and conclusions}\label{s:conclusions}

To conclude, the point spread function of an interferometer is integral to every observation but can have far-reaching effects when we are attempting to extract accurate flux density measurements. Whilst the effect of the PSF and CLEAN-based deconvolution methods have been noted for many decades \citep[e.g.,][]{1995Jorsater}, there have been only a few considerations in the literature \citep[e.g.,][]{benisty2021,Heywood2022}.

The flux density systematic error shown in this paper affects \textit{all} interferometric images as long as CLEAN-style deconvolution methods are still employed. While many previous interferometric arrays (e.g., the VLA) were originally designed to reduce the irregularities / non-Gaussianity of the PSF response, modern and upcoming interferometers (e.g., MeerKAT, LOFAR, SKA and the next-generation VLA), with their core dominated array configuration will naturally produce irregular PSFs. For example, in Figure~\ref{fig:meerkat_example} where we have simulated a 12-hour, 64-antenna MeerKAT observation at $\delta = -30^\circ$. The left-hand panel shows the resultant PSF which shows a widening due to the large number of short baselines present in the MeerKAT array. This is more clearly shown in the 1D slice of the PSF in the right-hand panel. 

It is likely that CLEAN-based deconvolution algorithms will still be used for many years to come due to their speed and versatility, therefore this effect should be taken into account when analysing any interferometric images. This means that users will need to adopt some of the solutions that were presented in Section~\ref{s:solns} and include these in their data analysis/reduction pipelines. However, the methods presented in this paper are not a comprehensive encyclopedia of all possible methods that could alleviate this effect but instead, we wish to open a discussion on the matter and ensure that this systematic does not need to be revisited in the future.

To mitigate this issue in its entirety, we will likely need to deviate from CLEAN-based deconvolution methods and adopt the more statistically robust algorithms that can produce more physically meaningful sky models \citep[e.g.,][]{Carrillo2012,arras2021}; however, it is still to be proven how these algorithms could be implemented on the datasets of the size that the next generation of interferometers will produce.

\section*{Acknowledgements}

We would like to extend our gratitude to the referee Prof. O. Smirnov and the scientific editor Prof. T. Pearson who helped improve this paper. We acknowledge the use of the ilifu cloud computing facility - \href{www.ilifu.ac.za}{www.ilifu.ac.za}, a partnership between the University of Cape Town, the University of the Western Cape, the University of Stellenbosch, Sol Plaatje University, the Cape Peninsula University of Technology and the South African Radio Astronomy Observatory. The ilifu facility is supported by contributions from the Inter-University Institute for Data Intensive Astronomy (IDIA - a partnership between the University of Cape Town, the University of Pretoria and the University of the Western Cape), the Computational Biology division at UCT and the Data Intensive Research Initiative of South Africa (DIRISA). We also acknowledge the Jodrell Bank Centre for Astrophysics, which is funded by the STFC. R.J.B. acknowledges funding from the European Union’s Horizon 2020 research and innovation programme under grant agreement No 101004719 (Option-RadioNet PILOT). The RADIOBLOCKS project will receive funding from the European Union’s Horizon Europe research and innovation programme under grant agreement No 101093934. A.N. has been supported by the Development in Africa with Radio Astronomy (DARA) project funded by STFC. $e$-MERLIN and formerly, MERLIN, is a National Facility operated by the University of Manchester at Jodrell Bank Observatory on behalf of STFC. This research makes use of \textsc{python} and the following packages: \textsc{matplotlib} \citep{matplotlib1,matplotlib2}, \textsc{astropy} \citep{astropy2013,astropy2018}, \textsc{numpy} \citep{numpy1,numpy2}, and \textsc{scipy} \citep{scipy}.

\section*{Data Availability}

The data underlying this article will be shared upon reasonable request to the corresponding author.

\bibliographystyle{mnras}
\bibliography{PSF} 

\bsp{}
\label{lastpage}
\end{document}